\documentclass[aps,prr,twocolumn,a4,longbibliography]{revtex4-1} 
\pdfoutput=1
\usepackage{amsmath}
\usepackage[dvipsnames]{xcolor}
\usepackage{graphicx}
\usepackage{setspace}
\usepackage{comment}
\usepackage{hyperref}
\usepackage{float}
\usepackage[normalem]{ulem} 
\usepackage{upgreek}

\newcommand{\STO}{SrTiO$_3$}
\newcommand{\STOO}{SrTiO$_3$ }
\newcommand{\STOov}{SrTiO$_{3-\delta}$}
\newcommand{\STOovv}{SrTiO$_{3-\delta}$ }
\newcommand{\tg}{$t_{2g}$}
\newcommand{\tgg}{$t_{2g}$ }
\newcommand{\egg}{$e_g$ }
\newcommand{\ef}{E$_{\rm F}$}
\newcommand{\ov}{O$_{\rm V}$}
\newcommand{\ovv}{O$_{\rm V}$ }
\newcommand{\uov}{$U$(O$_{\rm V}$)}
\newcommand{\uovv}{$U$(O$_{\rm V}$) }
\newcommand{\uti}{$U$(Ti)}
\newcommand{\utii}{$U$(Ti) }

\begin{document}

\title{Oxygen Vacancies in Strontium Titanate: a DFT+DMFT study}

\author{Jaime Souto-Casares}
\email{jaime.soutocasares@mat.ethz.ch}
\affiliation{Materials Theory, ETH Z\"urich, Wolfgang-Pauli-Strasse
  27, 8093 Z\"urich, Switzerland}
\author{Nicola A. Spaldin}
\email{nicola.spaldin@mat.ethz.ch}
\affiliation{Materials Theory, ETH Z\"urich, Wolfgang-Pauli-Strasse
  27, 8093 Z\"urich, Switzerland}
\author{Claude Ederer}
\email{claude.ederer@mat.ethz.ch}
\affiliation{Materials Theory, ETH Z\"urich, Wolfgang-Pauli-Strasse
  27, 8093 Z\"urich, Switzerland}

\date{\today}
\begin{abstract}

We address the long-standing question of the nature of oxygen vacancies in strontium titanate, using a combination of density functional theory and dynamical mean-field theory (DFT+DMFT) to investigate in particular the effect of vacancy-site correlations on the electronic properties. Our approach uses a minimal low-energy electronic subspace including the Ti-\tgg orbitals plus an additional vacancy-centered Wannier function, and provides an intuitive and physically transparent framework to study the effect of the local electron-electron interactions on the excess charge introduced by the oxygen vacancies. We estimate the strength of the screened interaction parameters using the constrained random phase approximation and find a sizeable Hubbard $U$ parameter for the vacancy orbital. Our main finding, which reconciles previous experimental and computational results, is that the ground state is {\it either} a state with double occupation of the localized defect state {\it or} a state with a singly-occupied vacancy and one electron transferred to the conduction band. The balance between these two competing states is determined by the strength of the interaction both on the vacancy and the Ti sites, and on the Ti-Ti distance across the vacancy. Finally, we contrast the case of vacancy doping in \STOO with doping via La substitution, and show that the latter is well described by a simple rigid-band picture.

\end{abstract}

\maketitle

\section{Introduction}
\label{intro}
Strontium titanate, \STO, is a perovskite-structure oxide with the ideal cubic $Pm\bar{3}m$ structure at room temperature, and a band-insulating electronic structure due to the formal $3d^0$ configuration of the Ti$^{4+}$ cations. In spite of its apparent simplicity, \STOO shows a wealth of interesting and sometimes technologically relevant properties, such as tunability of its high dielectric constant,\cite{fuchs99,hao_tunable06} quantum paraelectricity,\cite{barret52,mueller79} and even superconductivity.\cite{schooley64} Although these properties have been known for many years, there remain many open questions \cite{Collignon_et_al:2019}. For example, superconductivity occurs at unusually low doping levels \cite{Lin_et_al:2013}, indicating an exotic mechanism possibly related to its quantum paraelectricity.\cite{edge15} At the same time, reports of a two-dimensional electron gas\cite{ohtomo04} and emergent magnetism\cite{brinkman07} at surfaces and interfaces have rekindled interest in \STOO thin films for oxide electronics. 

All of these phenomena require the existence of electronic charge carriers, which are usually introduced through oxygen vacancies (\ov),\cite{eckstein07}  substitution of Sr$^{2+}$ by a trivalent ion such as La$^{3+}$, or of Ti$^{4+}$ by a pentavalent ion such as Nb$^{5+}$. Interestingly, the resulting properties can be quite sensitive to the specific type of doping.\cite{sarma96,wunderlich09,kinaci10,aiura19,higuchi00,ohta05,baniecki13,tokura93,sarma98,fujimori92,fujimori96} Regarding oxygen-vacancy doping in \STO, perhaps the most pressing open question is the nature of the introduced charge, with different experimental measurements leading to apparently contradictory  conclusions. On one hand, it is known that even very low concentrations of oxygen vacancies cause metallicity, with transport measurements indicating an increase in carrier density with increasing oxygen vacancy concentration.~\cite{tufte67,moos97,ohtomo_hwang07} On the other hand, there are multiple reports of optical absorption signals within the band gap,~\cite{leonelli1986,hasegawa2000,kan2005,yamada2009} such as for example a red luminescence at 2.0 eV, attributed to localized electrons forming Ti$^{3+}$ polarons trapped at isolated oxygen vacancies.\cite{crespillo18}

Computational studies aiming to clarify the physics of the \ovv state also show a range of conflicting scenarios, largely due to their different treatments of exchange and correlation effects. (For a detailed summary see Ref.~\onlinecite{hou10}.) Standard density functional calculations using the local density approximation (LDA) or generalized gradient approximation (GGA) predict a delocalized defect state at the bottom of the Ti-\tgg conduction band, consistent with the observed metallicity.~\cite{cuong2007,evarestov12,tanaka03} Use of B3PW hybrid functionals or the LDA+$\,{\cal U}$ method, however, lead to a doubly occupied in-gap state, $0.77$ eV (B3PW) or $0.11$ eV (LDA+$\,{\cal U}$, with $\mathcal{U} = 5$ eV) below the minimum of the conduction band, but do not capture the reported metallic conduction (although this can be reconciled by considering the formation of polarons).~\cite{franchini2015} Using spin-polarized GGA+$\,{\cal U}$ calculations with $\mathcal{U} = 5.0$ eV, the two electrons released by the missing oxygen are found to distribute between a localized magnetic in-gap state and a delocalized state in the conduction band, consistent with both sets of reported experimental behaviors.~\cite{hou10}  Finally, we mention a recent DFT$+\,\mathcal{U}+\mathcal{V}$ study, including self-consistent on-site and inter-site electronic interactions, which gives a good description of both stoichiometric and oxygen-deficient STO, with the details of the band structure and the vacancy formation energies agreeing well with experiments.\cite{ricca20} The authors also present a systematic study of the effects of various parameters such as structure and cell size, the used exchange-correlation functional, and the treatment of structural relaxations and spin polarization on the resulting properties.

These previous studies clearly indicate that electron interaction effects are important in describing the behavior of oxygen vacancies in SrTiO$_3$,~\footnote{The DFT+${\cal U}$ treatment is applied to the Ti-$d$ shells, so only the small fraction of the \ovv weight that might get projected onto the $d$ orbitals, whatever the formalism chosen, will actually feel the localization effect.} in spite of the band-insulating nature of the stoichiometric host material. However, while such DFT+$\,\mathcal{U}\, (+\mathcal{V})$ calculations are able to appropriately treat strong interactions between electrons in the transition-metal $d$ states (and in principle the oxygen $p$ states), this is usually achieved by introducing an artificial symmetry breaking resulting in long-range magnetic order. Furthermore, current implementations of the DFT+$\,\mathcal{U}$ formalism are typically based on projections on site-centered atomic orbitals, and so leave electrons at the vacancy site uncorrelated. The importance of explicitly considering interactions on the vacancy site was shown in model calculations using a minimal three-orbital model representing the vacancy, coupled to a bath representing the \tgg bulk bands,\cite{PhysRevLett.111.217601} but a full first-principles description is still lacking. 

Here, we study the electronic properties of oxygen-deficient \STOO using a combination of density functional theory (DFT) and dynamical mean-field theory (DMFT). The latter accounts for all dynamic correlation effects between electrons on the same ``site'' and thus introduces genuine many-body effects in the electronic structure obtained from DFT, and has already been applied recently to study oxygen vacancy complexes at the \STOO surface.~\cite{Lechermann_et_al:2016}
Similar to our previous work on oxygen-deficient LaTiO$_3$,\cite{souto19} we focus on the most important bands at the bottom of the conduction band, which can be expressed in a basis of maximally localized Wannier functions (MLWFs)~\cite{Marzari_et_al:2012,mostofi2014updated} with predominant Ti-\tgg character plus one additional Wannier function located at the vacancy site. This results in a physically intuitive framework to study the distribution of charge between the vacancy site and the Ti-\tgg conduction bands. 

We estimate the strength of the screened electron-electron interaction, both for the Ti-\tgg orbitals and for the vacancy level, using the constrained random phase approximation (cRPA).~\cite{aryasetiawan04,miyake2008,miyake2009,vaugier2012} We obtain a sizeable $U$ parameter for the vacancy level, confirming that the corresponding correlation effects should not be ignored. We then find that the occupation of the vacancy site is indeed controlled by the strength of the electron-electron repulsion both on the \ovv as well as on the Ti sites, balancing the system between a doubly-occupied localized vacancy state and a singly-occupied vacancy with the other electron doping the conduction band. The latter case can also be viewed as a site-selective Mott insulator, in which the singly occupied in-gap state corresponds to the lower Hubbard band on the vacancy site.

\begin{figure}
\includegraphics[width=0.45\textwidth]{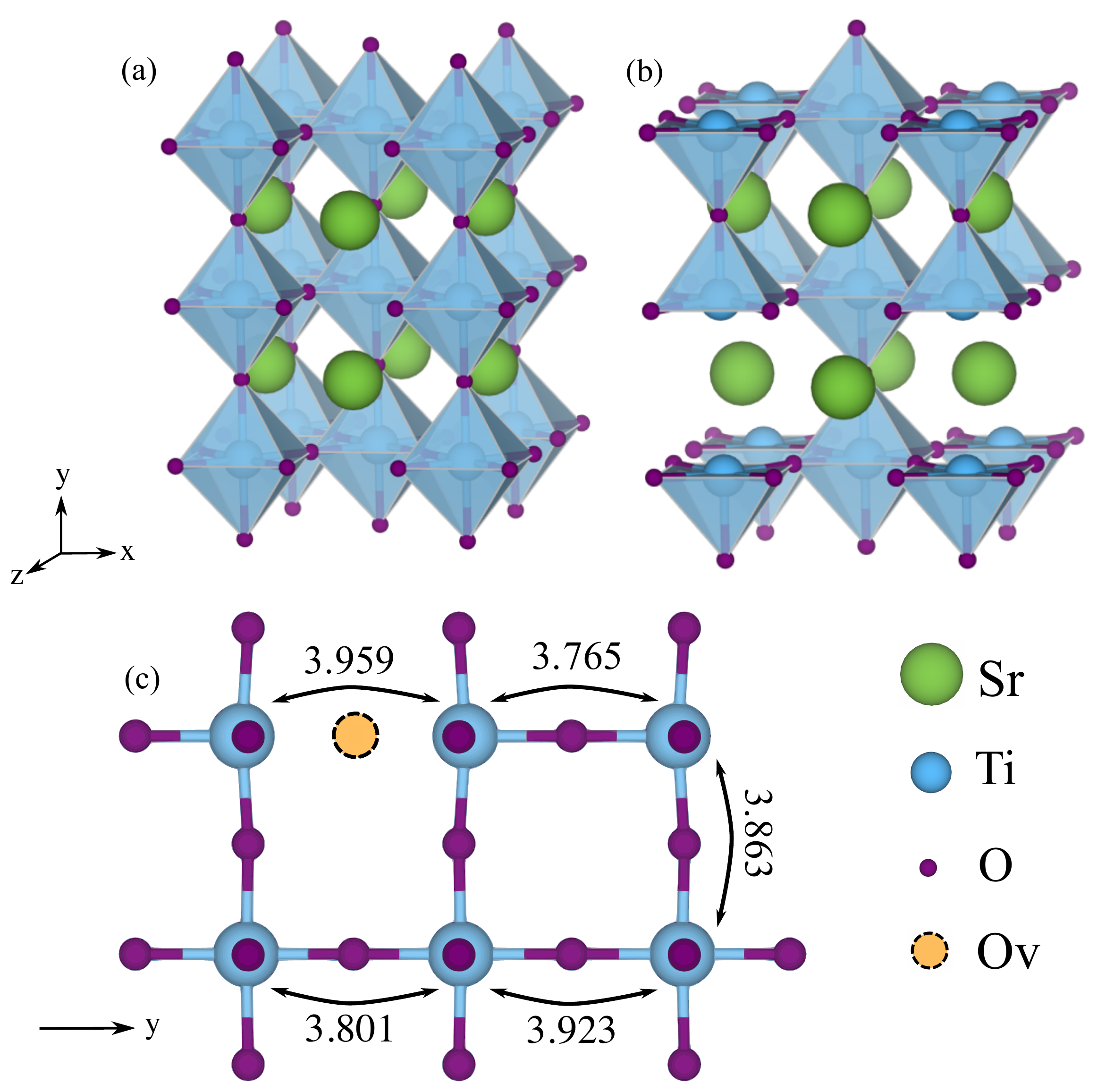}
\caption{(Color online) Supercells of the calculated structures for (a) \STOO (20-atom unit cell, {\it Pm$\bar{3}$m} symmetry) and (b) \STOov  (19-atom unit cell,  {\it P4/mmm} symmetry). (c) Geometry of the relaxed structure containing an oxygen vacancy (indicated with the orange circle) with the Ti-Ti distances given in \AA. For comparison, the calculated Ti-Ti distance for stoichiometric \STOO is $3.862$ \AA. Note that the orientation of part (c) is rotated relative to that of parts (a) and (b).}
\label{fig:geometries}
\end{figure}

\section{Computational Method}
\label{method}

The key point in our analysis is the explicit inclusion of an electronic orbital corresponding to the oxygen vacancy point defect into the low-energy Hamiltonian constructed from the DFT results, together with the usual Ti-\tgg bands. This allows us to explore in detail the effect of the electronic correlations on the vacancy site, and is the main novelty of the methodology presented here.

To obtain accurate geometries and initial bandstructures, we perform standard DFT calculations using the projector-augmented wave (PAW) method, as implemented in the ``Vienna ab-initio simulation package'' (VASP)~\cite{Kresse/Furthmueller_CMS:1996,Kresse/Joubert:1999}, version 5.4.1, together with the LDA exchange-correlation functional~\cite{LDA}. The valence configurations of the PAW potentials used are Sr($3s^23p^64s^2$), Ti($3s^23p^64s^13d^3$) and O($2s^22p^4$). To allow the system to accommodate the defect with a reasonable \ovv concentration, we use a $20$-atom unit cell, $19$-atom for the \ov-defective system SrTiO$_{2.75}$, corresponding to an oxygen vacancy concentration of 8.3$\%$.\footnote{This O$_{\rm V}$ concentration would correspond to an electron density slightly larger than typical highest values observed experimentally\cite{ohtomo_hwang07}, (5$\%$, but it should be noted that they assumed two electrons per defect). Nevertheless, it gives a reasonable compromise between computational efficiency and a realistic model.} Our calculations for La$_x$Sr$_{1-x}$TiO$_3$ (Section~\ref{ladoping}) are performed using larger 40-atom supercells. For calculations of geometries and band structures (Sections~\ref{geometry} and \ref{bandstructures}), well-converged results are obtained by sampling the Brillouin zone with a $8\times 8\times 8$ $\Gamma$-centered k-mesh and using a plane-wave energy cutoff of $800$ eV. For the more demanding cRPA calculations (Section~\ref{cRPA}), the k-space sampling is done with a $4\times 4\times 4$ mesh, and an energy cutoff of $500$ eV is used. In Section~\ref{dtiovti} (Ti-\ov-Ti distance dependence) these parameters are $8\times 8\times 8$ and $750$ eV, and in Section~\ref{ladoping} (comparison with La doping), these are $4\times 4\times 4$ and $700$ eV. Where lattice relaxation is employed, structural degrees of freedom are relaxed until forces fall below a $10^{-4}$ eV/\AA$\,$ threshold, with the symmetry constrained to disable the rotations of the oxygen octahedra that occur in SrTiO$_3$ below $\sim$100K. All calculations are performed with spin polarization excluded.

The low-energy correlated subspace for the DMFT calculations is then constructed using a basis of maximally localized Wannier functions (MLWF)~\cite{Marzari_et_al:2012,lechermann2006dynamical}, employing the Wannier90 code~\cite{mostofi2014updated}. We use the TRIQS/DFTTools package~\cite{PARCOLLET2015398,TRIQS/DFTTools}\footnote{A set of scripts and tools  can be found in \texttt{https://github.com/materialstheory/soliDMFT} .} to implement the DMFT calculations, averaging over both spin channels to enforce a paramagnetic solution. An effective impurity problem is solved for each inequivalent Ti site plus the vacancy site using the TRIQS/CTHYB solver~\cite{Seth2016274}, while the different impurity problems are coupled through the DMFT self-consistency. The \ovv site is treated at the same level as the Ti sites, as introduced in Ref.~\onlinecite{souto19} (more details about this implementation are given in Section~\ref{dmft}). The DFT+DMFT calculations are performed without full charge self-consistency. The local interaction is modeled using the Hubbard-Kanamori parametrization with spin-flip and pair-hopping terms included.~\cite{Castellani/Natoli/Ranninger:1978} Within the Hubbard-Kanamori parametrization, the strength of the electron-electron interaction is described by the intra-orbital Hubbard parameter $U$ and the Hund coupling parameter $J$. Note that these parameters are different from the average interaction parameters ${\cal U}$ and ${\cal J}$ typically used in DFT+$\,{\cal U}$ calculations. The double counting correction is computed within the fully localized limit according to Held\cite{held2007electronic}, and all calculations are performed at room temperature, $\beta=(k_BT)^{-1}=40$ eV$^{-1}$. We use a fixed value of $J=0.64$ eV on the Ti sites, whereas the values for the Hubbard $U$, both on the vacancy and the Ti sites, are varied to analyze the effect on the electronic properties. Full frequency spectral functions, $A(\omega)$, are obtained from the local Green's functions in imaginary time, $G(\tau)$, using the Maximum Entropy algorithm~\cite{Bryan1990}. The spectral weight around the Fermi energy, $\bar{A}(0)$, is calculated from the impurity Green's function as $\bar{A}(0)=-\beta/\pi\, G(\beta/2)$. The quasiparticle weight, $Z$, is calculated for each site as $Z=[1- \Sigma(i\omega_0)]^{-1}$, where $\Sigma(i\omega_0)$ is its self-energy at $\omega_0$, the smallest calculated Matsubara frequency. 

We calculate the screened Coulomb interaction within the low-energy correlated subspace using the constrained Random Phase Approximation (cRPA).~\cite{aryasetiawan04,miyake2008,miyake2009,vaugier2012}. In the cRPA method, a partial polarization function, $P_r$, is calculated by excluding all possible electronic transitions taking place within the correlated subspace. The bare Coulomb interaction, $v$, is then renormalized through screening by the higher energy degrees of freedom through this $P_r$, yielding the frequency-dependent partially screened Coulomb interaction $W_r(\omega) = [1-vP_r(\omega)]^{-1}v$. The local interaction parameters of the Kanamori Hamiltonian, $U$ and $J$, are then obtained from the static limit by calculating matrix elements of $W_r(\omega=0)$, with the MLWFs used as local basis orbitals within the DMFT calculation, and taking appropriate averages over orbitals.

\section{Results and Discussion}
\label{Results}

\subsection{Geometry optimization}
\label{geometry}

Our calculated optimized LDA lattice constant for \STOO constrained to cubic $Pm\bar{3}m$ symmetry is 3.862 \AA. (For a picture of the $20$-atom ($\sqrt 2,\,2,\sqrt 2$) unit see Fig.~\ref{fig:geometries}(a)).
This value agrees well with literature LDA calculations,\cite{tanaka03} and it is not too far from the experimental value of $3.900$ \AA$\,$.~\cite{cao00} In subsequent relaxations after removing one oxygen (Fig.~\ref{fig:geometries}(b)), we keep the lattice constants fixed to our calculated values for stoichiometric \STO.~\footnote{Even though the presence of O$_{\rm V}$ can cause an expansion with respect to the pure SrTiO$_3$, the volume of SrTiO$_{3-\delta}$ is kept fixed in order to prevent the interaction between periodic images of the unit cell.}
The resulting structure and Ti-Ti distances are shown in Fig.~\ref{fig:geometries}(c). The removal of one of the oxygen atoms lowers the symmetry of the formerly cubic
crystal to tetragonal {\it P4/mmm}, in which the unique axis corresponds to the Ti-\ov-Ti chain ($y$ axis in Fig.~\ref{fig:geometries}(b)),
and divides the Ti sites in two inequivalent types: one next to the vacancy, with five Ti-O bonds, and one farther from the vacancy, with a complete oxygen coordination octahedron.
(Note that we do not include the low-temperature antiferrodistortive rotations of the oxygen octahedra; our preliminary tests suggest that their effect on the vacancy state is rather weak.)
The Ti-Ti distances in the plane perpendicular to the tetragonal axis are slightly modified, with the vacancy pushing the two closest Ti apart, increasing their distance by $2.5 \%$ with respect to the distance in stoichiometric \STO. 
The next Ti-O-Ti distance on the same axis contracts by the same amount, due to the volume constraint. On the other inequivalent Ti-O-Ti chain along the $y$ axis there is also a $1.6 \%$ contraction (expansion) of the Ti-O-Ti distance at the same (different) $y$ position as the closest Ti-\ov-Ti.

\subsection{DFT Bandstructures}
\label{bandstructures}

\begin{figure}
\includegraphics[width=0.45\textwidth]{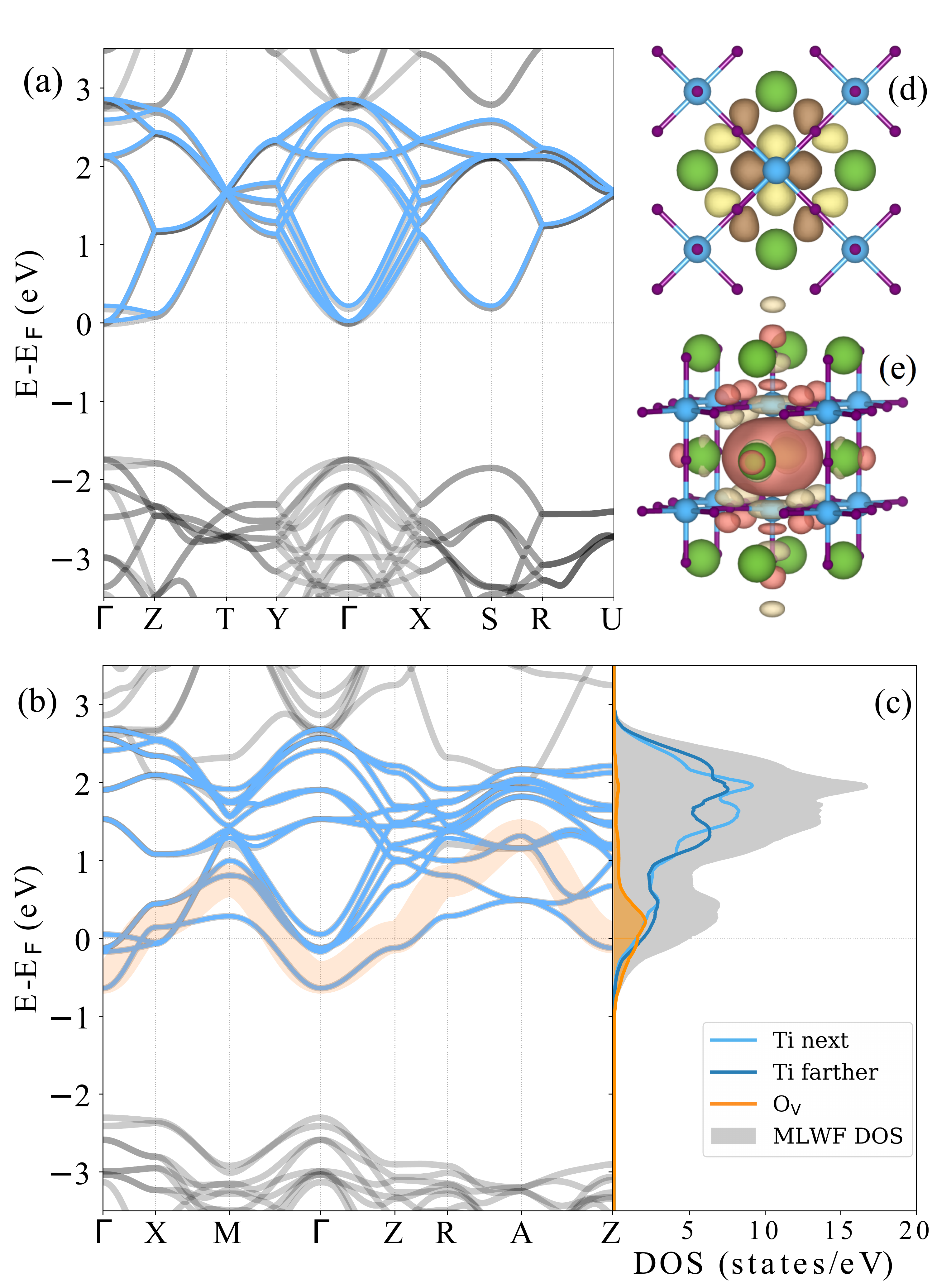}
\caption{(Color online) Calculated DFT bandstructures for (a) stoichiometric \STO , and (b) SrTiO$_{2.75}$. DFT and MLWF bands are shown as grey and blue solid lines, respectively; the vacancy band in (b) is highlighted in orange (see  main text). (c) MLWF-projected density of states (DOS) for the {\it t$_{2g}$}-like and \ovv MLWFs. (d) shows the real-space representation of a $d_{xz}$-type MLWF in the stoichiometric \STOO system, while (e) corresponds to the \ovv MLWF.}
\label{fig:bands}
\end{figure}

The calculated bandstructures for stoichiometric \STOO and defective \STOovv are shown and compared in Fig.~\ref{fig:bands}. The low-energy region around the gap is composed of a valence band of mostly O-$p$ character, and a conduction band whose bottom part has mainly Ti-\tgg contributions, with some weight coming from the O-$p$ orbitals, and minimal overlap with the Ti-\egg and Sr-$s$ bands at $\sim\!\!3$ eV above the gap. Valence and conduction bands are separated by a gap of $1.8$ eV, strongly underestimating the experimental value of $3.25$ eV.~\cite{benthem01} 
The bands between approximately 0 and 3\,eV can be expressed in terms of MLWFs centered on the Ti atoms and showing a strong \tgg orbital character with additional admixtures of O-$p$ on the surrounding ligands, see Fig.\ref{fig:bands} (d).

The removal of one oxygen atom changes mainly the lower part of the conduction band region of the bandstructure, with the appearance of an additional band that crosses the Fermi energy, \ef, making the system metallic, and accompanied by a lifting of degeneracies at the special points of the Brillouin zone. This band could accommodate, in principle, the two electrons released by the vacancy. However, its overlap with the bottom of the conduction band edge (Fig.~\ref{fig:bands}) results in a partial transfer of charge into the Ti-\tgg bands. Therefore, a complete description of the low-energy behaviour of the system must include both the Ti-\tgg bands and this \ov-induced band, and we proceed by incorporating it into the minimal basis of Ti \tg-like MLWFs in the DMFT treatment. We thus construct $12+1$ MLWFs, using initial \tgg projections on the Ti sites plus one $s$-like projection centered around the vacancy site.

The relationship of this new band to the \ov\ is clearly seen from the MLWF centered on the position of the missing oxygen: its real-space representation shows an approximately spherical orbital centered around the vacancy with tails reaching to the neighbouring ions (Fig.~\ref{fig:bands}(e)). Moreover, if one extracts a single MLWF centered on the vacancy site, the resulting Wannier band closely follows the new \ov-induced Bloch band (orange thick line in Fig.~\ref{fig:bands}(b)). As one can see from the MLWF-projected density of states (Fig.~\ref{fig:bands}(c)), the \ov-centered  MLWF accounts for most of the weight of this additional band. The occupations of the Wannier states, as calculated in DFT, are $0.21$, $0.28$, and $1.02$ for the two inequivalent Ti sites, next to and farther from the \ov, and the \ovv site itself, respectively, in units of the electron charge.

\subsection{DMFT results}
\label{dmft}

\begin{figure*}
\includegraphics[width=0.9\textwidth]{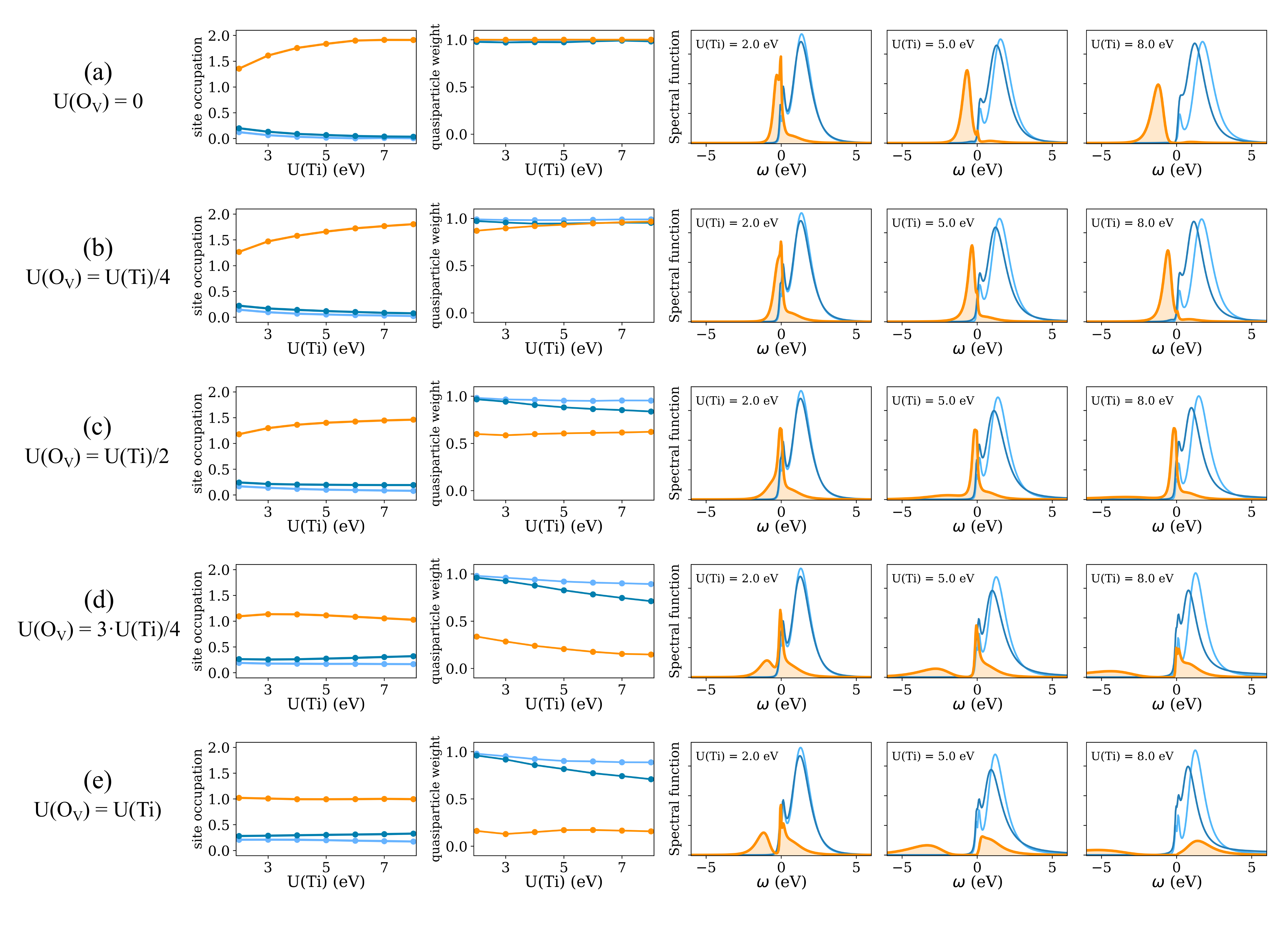}
\caption{(Color online) DFT+DMFT results for different settings of U(\ov)/U(Ti), from $0$ (a) to $1$ (e). The left column in each row shows the site occupations for the three types of correlated sites: Ti-next, Ti-farther, and O$_{\rm V}$ (lighter blue, darker blue and orange, respectively). The second to left column shows the corresponding quasiparticle weights $Z$. The following three plots in each row show the spectral functions for the three aforementioned sites for three different \utii values.
}
\label{fig:dmft}
\end{figure*}

Next, we perform DMFT calculations to investigate how an explicit local Hubbard-like interaction between the electrons affects the spectral properties and the charge distribution between the vacancy state and the Ti-\tgg bands.
The input for the DMFT calculations is constructed from the tight-binding-like Hamiltonian expressed in the basis of the MLWFs, plus the Coulomb matrix that models the electron-electron interaction. As stated in Sec.~\ref{method}, we simplify the latter for the \tgg states by using the Kanamori form, for which only two site-dependent parameters have to be specified: $U$, the on-site intra-orbital Hubbard repulsion, and $J$, the Hund's coupling. For the \ovv site with only one orbital, there is only one parameter, $U$, describing the corresponding intra-orbital Coulomb repulsion. This procedure allows us to independently vary the strength of the local interaction on the different types of sites, including the vacancy, and hence to determine the evolution of the system when \uovv changes independently of \uti. 
We note that for other early transition metal perovskites, such as, e.g. LaTiO$_3$, values of \utii between 4-5\,eV have often led to good agreement with experimental observations when using a minimal \tgg orbital subspace;~\cite{pavarini2004,dymkowski2014} we therefore choose the range from 2-8\,eV. On the other hand, given the lack of chemical intuition for choosing \uov, we have considered \uov$=0$ and \uov$=$\utii as reasonable limits.

Fig.~\ref{fig:dmft} shows our calculated DMFT site occupations, along with the site-resolved quasiparticle weight $Z$, and the corresponding spectral functions, for different choices of $U$ on the Ti and the \ovv sites. The top row (Fig.~\ref{fig:dmft}(a)) corresponds to \uov$=0$, that is treating the electrons on the vacancy site as ``uncorrelated''. In this case, the effect of increasing \utii is to increase the \ovv occupation from $<1.5$ for small \utii to nearly 2, that is almost completely filled, for large \uti. For \utii$>6$ eV, a metal-insulator-transition (MIT) is observed, in which the Ti bands are totally depleted, and the system becomes a band insulator. This MIT is related to a shift of the \ovv (Ti) spectral weight to lower (higher) energies, until the corresponding overlap vanishes (see spectral function for \utii$=8$ eV). 
The quasiparticle weight $Z$ is close to 1 for all sites, independent of \uti, indicating that the electrons remain uncorrelated, consistent with the transition from an uncorrelated metal to a band insulator.~\footnote{Note that in the insulating state the quantity we plot, $[1-\Sigma(i\omega_0)]^{-1}$, loses its meaning as approximate quasiparticle weight $Z$. Nevertheless, we include the corresponding data in our plots for consistency.} Our DFT+DMFT result for higher \utii and \uov$=0$ is thus equivalent to the results obtained in previous studies within DFT+$\,\mathcal{U}$,\cite{cuong2007} as discussed in Sec.~\ref{intro}.

The remaining rows in Fig.~\ref{fig:dmft}, (b)-(e), demonstrate the effect of introducing and then increasing a local Coulomb repulsion on the vacancy site, with \uov$\leq$\uti. 
By following the evolution of the \ovv site occupation as \uov$/$\utii grows, we can see that the main effect of \uovv is to first weaken the effect of \utii towards establishing a doubly-occupied vacancy site (see cases with $0 <$ \uov $\leq$ \uti$/2$ in Fig.~\ref{fig:dmft}(b) and (c)), and then, for \uov$>$\uti/2, to drive the system instead towards a state with a half-filled \ovv and one electron doped into the Ti bands, see Figs.~\ref{fig:dmft}(d) and (e). In this limit, the Ti sites obtain an average filling of about $0.25$, with a slightly higher occupation of the Ti farther away from the vacancy.

Increasing \uovv also strongly reduces the quasiparticle weight on the vacancy site, indicating strong local electronic correlations. Simultaneously, \uovv triggers a reduction of $Z$ on the Ti sites, in particular on the site farther away from the vacancy. One can also observe a pronounced effect on the site-resolved spectral functions. In particular, increasing \uovv results in a clear gap opening in the spectral function on the vacancy site once it reaches half-filling. This indicates a site-selective Mott transition, that is a localization of one electron on the vacancy site while the doped Ti bands remain metallic, albeit with a quasiparticle renormalization on the farther Ti site of $Z \approx 0.7$.
Thus, this regime is characterized by a metallic conduction band doped with one electron per vacancy, and a split-off ``in-gap'' state containing one electron localized on the vacancy site. In the picture of the site-selective Mott insulator, this in-gap state corresponds to the lower Hubbard band of the vacancy site spectral function.
Note that, in this regime of \uov$=$\utii the filled region of the \ovv spectral function overlaps with the O-$p$ bands, which start $2$ eV below $E_{\rm F}$, but are excluded from the DMFT calculation.

Our results show that the specific treatment of electronic correlations on the \ovv site has a strong influence on whether the two electrons released by the missing oxygen are itinerant or localized. Several scenarios that have previously been discussed and reported for \STOovv can be realized with particular choices of \utii and \uov, from the band-insulating limit with localization of the vacancy charge in the band gap~\cite{alexandrov2009} (high \uti, zero \uov) to the case of a paramagnetic impurity, in which one electron remains trapped in the gap while the other is delocalized into the Ti $d$-bands (\uti$=$\uov).~\cite{hou10} We note that the lack of charge self-consistency in our calculations may diminish the quantitative nature of these results; however, the qualitative picture should remain valid.

\subsection{cRPA calculations}
\label{cRPA}

The results discussed in the previous section show that the localization/delocalization of the two electrons associated with the missing oxygen depends critically on the values of the interaction parameters, \utii and \uov. In some cases, values for the interaction parameters can be estimated from previous experience or from comparison of certain calculated quantities with corresponding experimental measurements. In the present case, however, there is a lack of consensus among experimental studies (as described in Sec.~\ref{intro}). Furthermore, due to the use of a non-standard orbital basis, in particular on the vacancy site, it is not {\it a priori} clear what values for \utii and \uovv will provide the most realistic description of \STOov.  
Therefore, we now estimate the strength of the screened Coulomb repulsion corresponding to our orbital basis using the constrained random phase approximation (cRPA).~\cite{aryasetiawan04,miyake2008,miyake2009,vaugier2012} 

\begin{figure}
\includegraphics[width=0.35\textwidth]{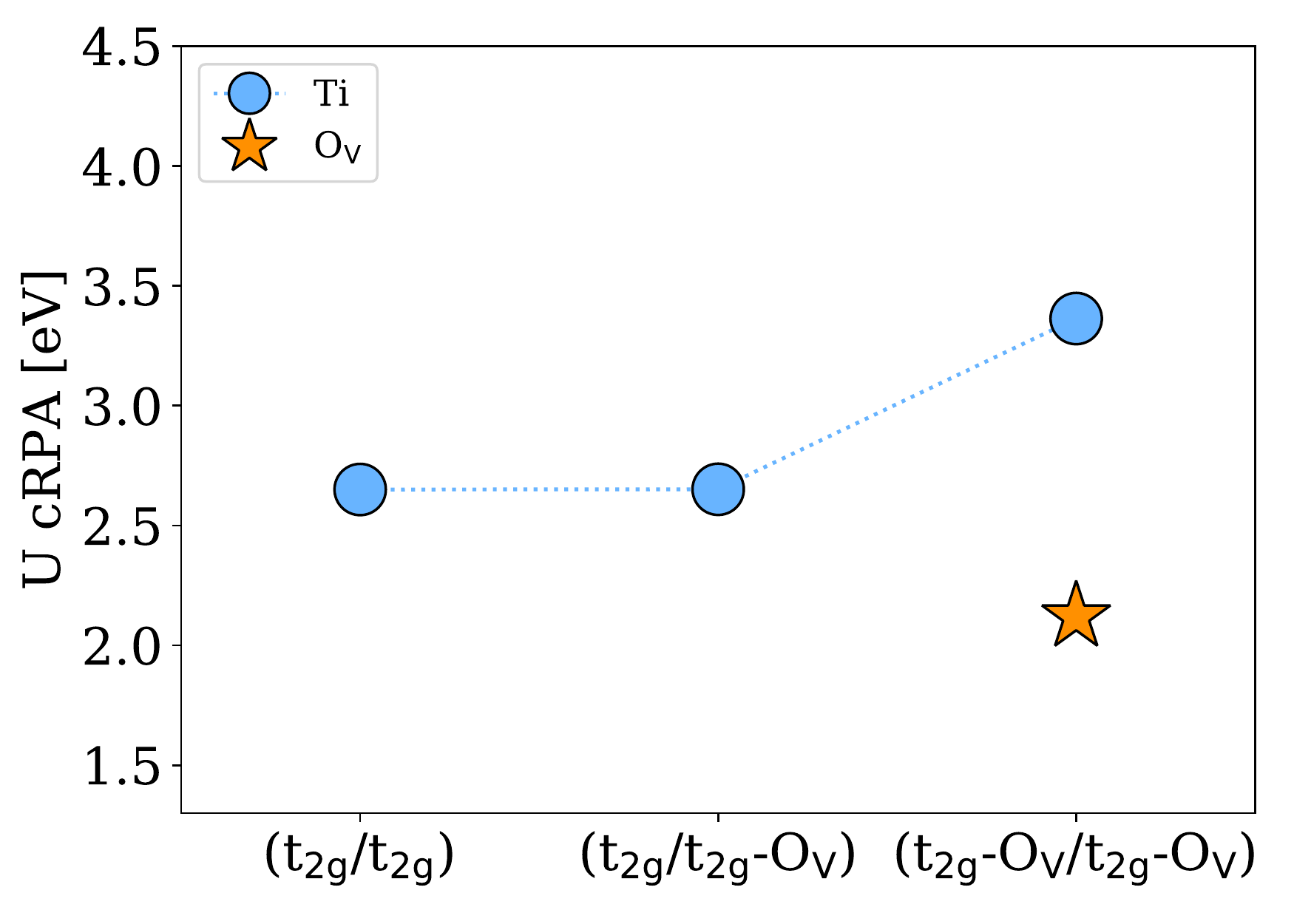}
\caption{(Color online) Averaged values for the partially screened interaction parameters \utii and \uovv obtained within cRPA. Blue dots represent the values for the Ti sites, while the orange star corresponds to the \ovv site. Displayed on the horizontal axis are the different schemes for choosing the target subspace (see the main text for more details).}
\label{fig:crpa}
\end{figure}

Within cRPA, the electronic degrees of freedom are divided into a ``screening subspace'' and a ``target subspace'', and excitations taking place exclusively within the target subspace are excluded from the screening (see Sec.~\ref{method}). 
Generally, the cRPA target subspace should be identical to the correlated subspace used for the DMFT calculation. However, there are also cases where a certain number of ``uncorrelated'' bands are included in the DMFT subspace, without considering a local Hubbard-like interaction for these bands. In such cases it might be appropriate to include these uncorrelated bands within the screening subspace in the cRPA calculation.~\cite{miyake2008,miyake2009} 

Thus, in order to gain additional insights, and to see how the calculated values depend on the specific sub-division of electronic degrees of freedom, we consider three different cases. 
In the first case, we construct only three \tg-like Wannier functions for each Ti site. These orbitals then also define the cRPA target subspace, achieved through a projection of the Bloch states onto these Wannier orbitals.
All other bands, including that related to the vacancy, act as the screening subspace. Following the notation established in Ref.~\onlinecite{miyake2008} and used in several other studies,\cite{vaugier2012,seth2017,amadon2014} we denote this case as (\tg/\tg). Here, the first symbol denotes the orbitals/bands defining the cRPA target subspace, while the second symbol indicates the whole set of Wannier functions that has been constructed. The screened interaction parameters are then evaluated for the Wannier functions spanning the cRPA target subspace.  

For the second scenario, (\tg{\it -O$_{V}$}/\tg{\it -O$_{V}$}), a full Wannier representation of the low energy conduction bands in \STOovv containing three \tg-like Wannier functions per Ti plus an additional one located on the vacancy site is constructed, as described in the previous section. This whole set of Wannier functions is then used to define the cRPA target subspace. Thus, in this case the vacancy band is excluded from the screening and included in the target/correlated subspace. 

The third, intermediate, case is denoted as (\tg/\tg{\it -O$_{V}$}). Here, the same Wannier functions as in the previous case are constructed, but only the subset of \tg-like functions are used to define the cRPA target subspace. This essentially means that the effective interaction parameters calculated for the Ti-\tgg orbitals also include screening processes involving the vacancy-band, similar to the first case, but using exactly the same Wannier representation as in the second scenario.

Fig.~\ref{fig:crpa} shows the values of the screened intra-orbital interaction parameters \utii and \uov, given by the corresponding averaged diagonal elements of the calculated $U$ tensor for all three cases.
The difference in \utii between the (\tg/\tg) and (\tg/\tg{\it -O$_{V}$}) cases is negligible, in spite of the fact that the average quadratic spread of the Ti \tgg{} Wannier orbitals is reduced from 3.3\,\AA$^2$ to 2.0\,\AA$^2$ (4.9\,\AA$^2$ for the O$_V$ orbital). However, this merely increases the bare (unscreeend) interaction parameter from 14.1\,eV to 14.9\,eV, indicating that the dominant factor in determining the final $U$ value is the screening.
Thus, removing the vacancy band from the screening channel has a stronger effect, as can be seen by comparing cases (\tg/\tg{\it -O$_{V}$}) and (\tg{\it -O$_{V}$}/\tg{\it -O$_{V}$}), resulting in an increase of \utii from 2.65\,eV to 3.47\,eV, and demonstrating the sensitivity of $U$ on the specific screening channel.
In addition, a sizeable interaction parameter of \uovv$ = 2.12$\,eV is obtained, which corresponds to approximately $60\%$ of \uti.
This clearly shows that interaction effects cannot be neglected for the vacancy orbital, and that, in addition, the presence of the vacancy band strongly influences the effective interaction parameters on the Ti sites.
These findings thus further support our approach of including an explicit treatment of correlation effects on the vacancy orbital, reinforcing the crucial role played by both \utii and \uovv already presented in Sec.~\ref{dmft}.

The values of the averaged interaction parameters obtained for the case (\tg{\it -O$_{V}$}/\tg{\it -O$_{V}$}), \utii $\approx 3.5$\,eV and \uovv $\approx 2.1$\,eV, would correspond to DFT+DMFT results with a vacancy occupation slightly larger than 1 and a corresponding quasiparticle weight below 0.5, but still with a metallic spectral function on the vacancy site (Fig.~\ref{fig:dmft}). 
We note, however, that simply using the static ($\omega=0$) value of the screened interaction in a DFT+DMFT calculation with frequency-independent local interaction, might underestimate the corresponding interaction effects.
Furthermore, recent work has shown that the random phase approximation can lead to an ``overscreening'' of the local interaction,~\cite{werner2018} and thus an underestimation of $U$, in particular for strongly correlated systems.~\cite{haule2018}
The cRPA results should thus rather be viewed as providing a rough ballpark, or a lower bound, rather than definite values to be used in realistic DFT+DMFT calculations.\footnote{Moreover, the $U$-regimes of the localization/delocalization boundary for the vacancy site might also change if charge self-consistency were included in the calculation.}

\subsection{Influence of the Ti-\ov-Ti distance}
\label{dtiovti}

As shown Fig.~\ref{fig:geometries} and discussed in Sec.~\ref{geometry}, structural relaxation using the standard LDA results in an outward relaxation of the two Ti atoms adjacent to the vacancy, and thus an elongation of the Ti-\ov-Ti distance across the vacancy compared to the Ti-O-Ti distance in the ideal stoichiometric case.
While this is consistent with previous work using LDA or GGA,~\cite{buban04,luo04} other calculations using hybrid functionals or LDA+$\mathcal{U}$ corrections have instead found a contraction of the Ti-\ov-Ti distance, independent of other factors such as the size of the supercell.~\cite{grayznov13,mitra12}
As mentioned previously (see Sec.~\ref{intro}), the latter calculations also find a tendency for charge localization, whereas LDA/GGA result in some charge spilling into the Ti bands. Therefore, we next manually tune the Ti-\ov-Ti distance and calculate the resulting changes in electronic properties.

\begin{figure}
\includegraphics[width=0.35\textwidth]{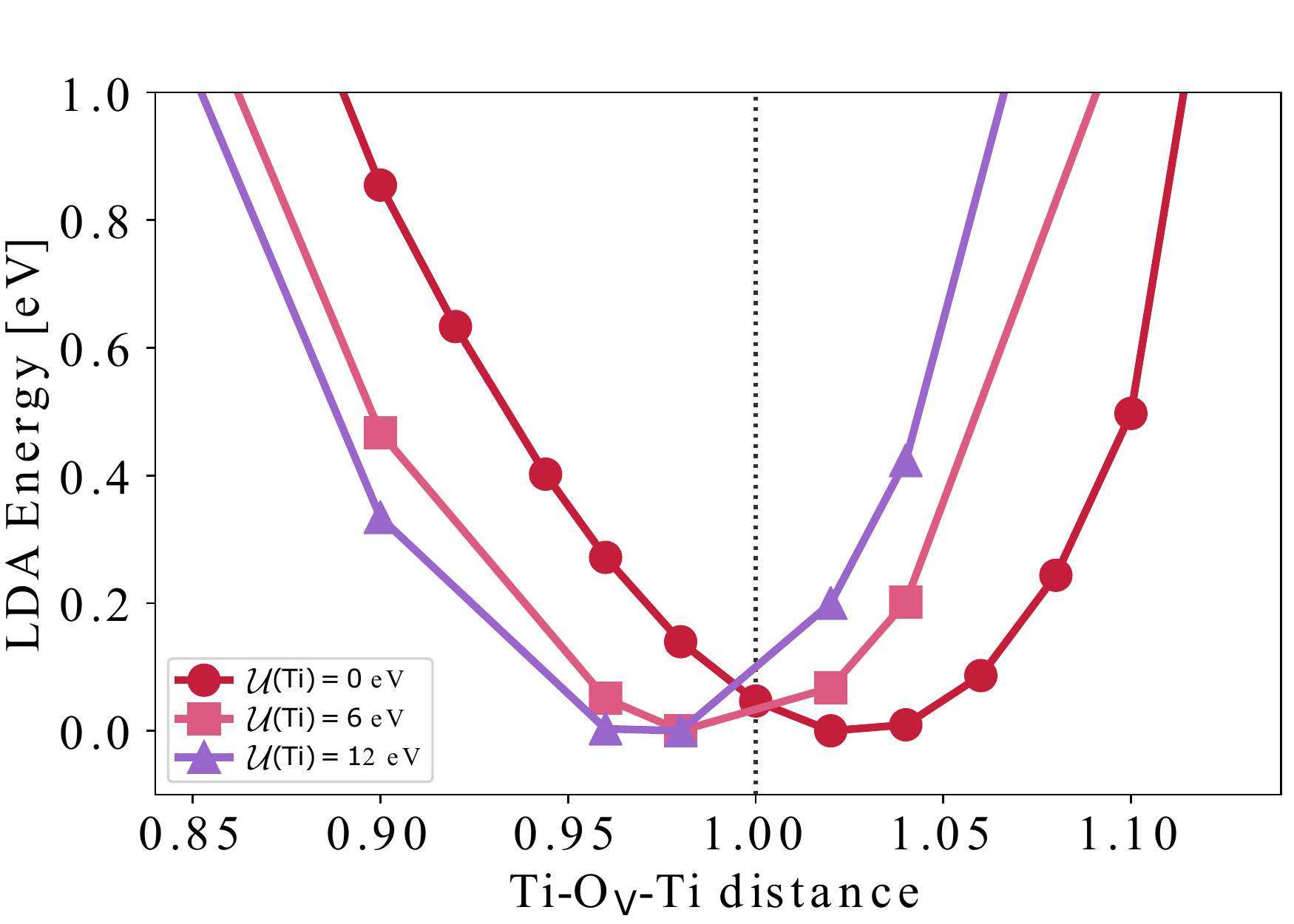}
\caption{(Color online) Total energy as function of the Ti-\ov-Ti distance (in units of the Ti-O-Ti distance in stoichiometric \STO) obtained within LDA+$\mathcal{U}$ for three different choices of the Hubbard $\mathcal{U}$ parameter. The vertical line highlights the distance in cubic stoichiometric \STOO}
\label{fig:energy_distance}
\end{figure}

Fig.~\ref{fig:energy_distance} shows the total energy calculated using LDA and LDA+$\,{\cal U}$ for 19-atom \STOovv as a function of the Ti-\ov-Ti distance (in units of the Ti-O-Ti distance in stoichiometric \STO). In these calculations, we only allow for a structural relaxation of those oxygen atoms that are situated next to the Ti in the same plane perpendicular to the Ti-\ov-Ti direction. Calculations are performed for three different choices of the static Hubbard $\mathcal{U}$ parameter.~\cite{dudarev1998}
The case with ${\cal U}=0$ corresponds to a standard LDA calculation and leads to an increase of the Ti-\ov-Ti distance, consistent with the results presented in Sec.~\ref{geometry}. Increasing $\mathcal{U}$ leads to a decrease of the relaxed Ti-\ov-Ti distance, in accordance with the aforementioned trend in the literature, ultimately shortening this distance with respect to that in pure \STO.

\begin{figure*}
\includegraphics[width=\textwidth]{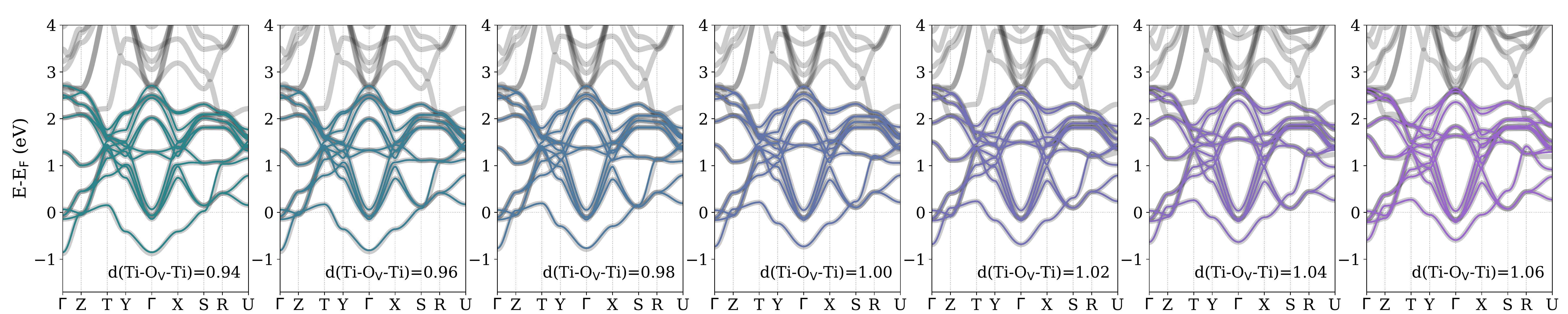}
\caption{(Color online) Evolution of the \STOovv bandstructure with respect to the Ti-\ov-Ti distance, measured in units of the Ti-O-Ti distance in stoichiometric \STO. The central plot corresponds to the Ti-O-Ti distance in stoichiometric \STO. We see that bringing the two Ti ions closer together (left side) pushes the \ovv band down in energy, favouring charge localization, while pulling the Ti atoms apart (right side) enhances the entanglement of the \ovv band with the Ti-\tgg bands, and redistributes some of its charge onto the Ti ions.}
\label{fig:bands_distance}
\end{figure*}

\begin{figure}
\includegraphics[width=0.4\textwidth]{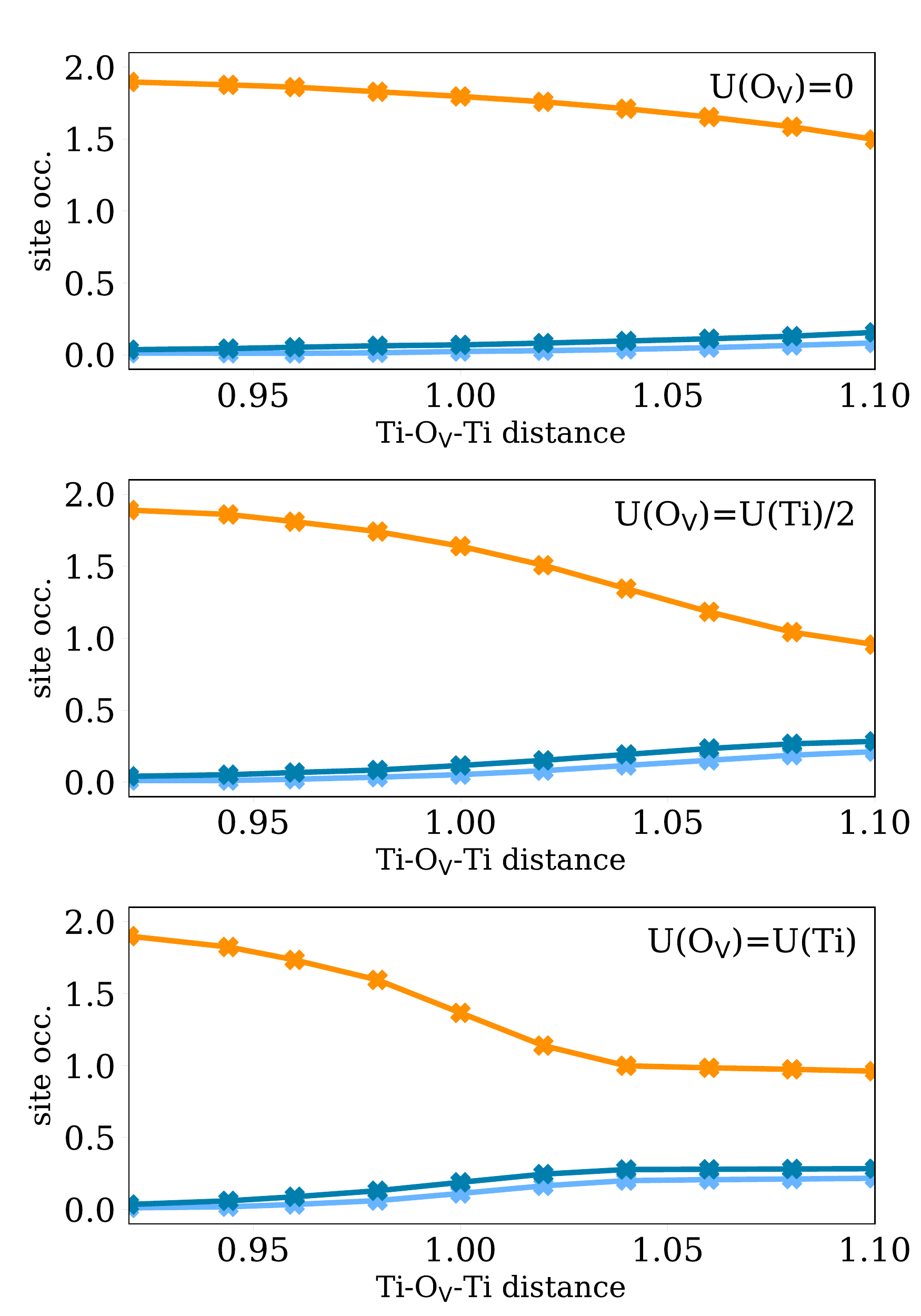}
\caption{(Color online) \STOovv DMFT occupations for the three different sites as a function of the Ti-\ov-Ti distance (in units of the Ti-O-Ti distance in stoichiometric \STO), for three different values of \uov$/$\uti). Lighter blue, darker blue, and orange represent Ti-next, Ti-farther, and O$_{\rm V}$ sites, respectively. Electron transfer from the \ovv to the Ti ions is favored by large Ti-\ov-Ti distance and by large $U$(\ov).}
\label{fig:dmft_distance}
\end{figure}

In order to further explore the relationship between the localization of the defect state and the Ti-\ov-Ti distance, we perform LDA calculations for a \STOovv structure in which we have systematically varied the Ti-\ov-Ti distance. For each (fixed) Ti-\ov-Ti distance, the oxygen ions are allowed to relax within the tetragonal symmetry constraint, while all other atoms and the lattice constant are kept fixed. 
Fig.~\ref{fig:bands_distance} shows the evolution of the resulting DFT bandstructure as the Ti-\ov-Ti distance is varied over a range that includes also the various equilibrium values obtained for different values of $\mathcal{U}$ within DFT+$\,{\cal U}$.
We see that the Ti-\ov-Ti distance affects primarily the position of the vacancy band, which is higher in energy, and therefore contains less electronic charge, at larger Ti-\ov-Ti distances. We conclude that short Ti-\ov-Ti distances tend to localize the charge around the defect, while longer distances tend to partially delocalize the charge. 

Following the same procedure as in Sec.~\ref{dmft}, we then perform a series of DMFT calculations for the different structures corresponding to different Ti-\ov-Ti distances. We use \uti $=4$ eV and three different values for \uov{} (0, \uti/2, and \uti). Our calculated site occupations are displayed in Fig.~\ref{fig:dmft_distance}. We see that for all values of \uov, the occupation of the vacancy is decreased with increasing Ti-\ov-Ti distance, while the Ti occupancies increase proportionally. The most pronounced change in occupation is observed for \uov$=$\utii (Fig.~\ref{fig:dmft_distance}, lower panel). In this case, the vacancy site occupation drops from close to $2$ (for the shortest considered Ti-\ov-Ti distance of $0.92$) down to $1$ (for distances equal or larger than $1.04$). This case corresponds to the site-selective Mott-insulating state obtained previously in Sec.~\ref{dmft}, in which the vacancy occupation is essentially locked to 1 due to the gap in the corresponding local spectral function. The strong dependence of the DMFT occupancies on \uovv for larger Ti-\ov-Ti distances is of course consistent with the results already presented in Fig.~\ref{fig:dmft}, where the structure was relaxed within LDA, leading to a Ti-\ov-Ti distance of $1.024$ in these units.

These results indicate that the site-selective Mott insulating state, corresponding to a singly occupied vacancy site, is favorable for an elongated Ti-\ov-Ti distance. We note that the lack of charge self-consistency does not allow us to extract reliable DFT+DMFT total energies as a function of the Ti-\ov-Ti distance.

\subsection{Comparison with La$_{\rm Sr}$ substitution}
\label{ladoping}

\begin{figure}
\includegraphics[width=0.5\textwidth]{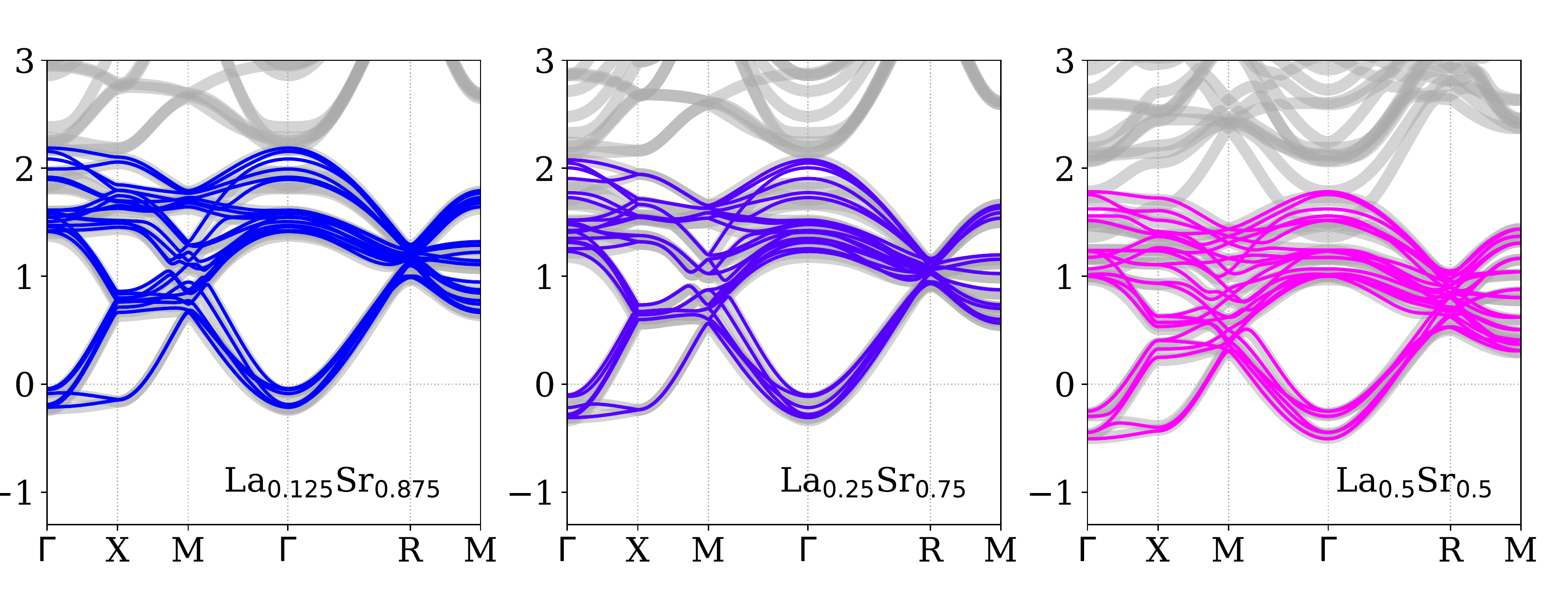}
\caption{(Color online) LDA Bandstructure of La$_x$Sr$_{1-x}$TiO$_3$ for different values of $x$. DFT and MLWF bands are shown in grey and in colour (from blue to red), respectively. While the general shape and bandwidth of the Ti-\tgg remains basically constant for the whole series, its relative position with respect to \ef{} moves gradually to lower energies for increasing $x$, marking the expected filling of the \tgg bands from $d^0$ for \STOO ($x=0$) to $d^1$ in LaTiO$_3$ ($x=1$).}
\label{fig:la_bands}
\end{figure}

\begin{figure}
\includegraphics[width=0.5\textwidth]{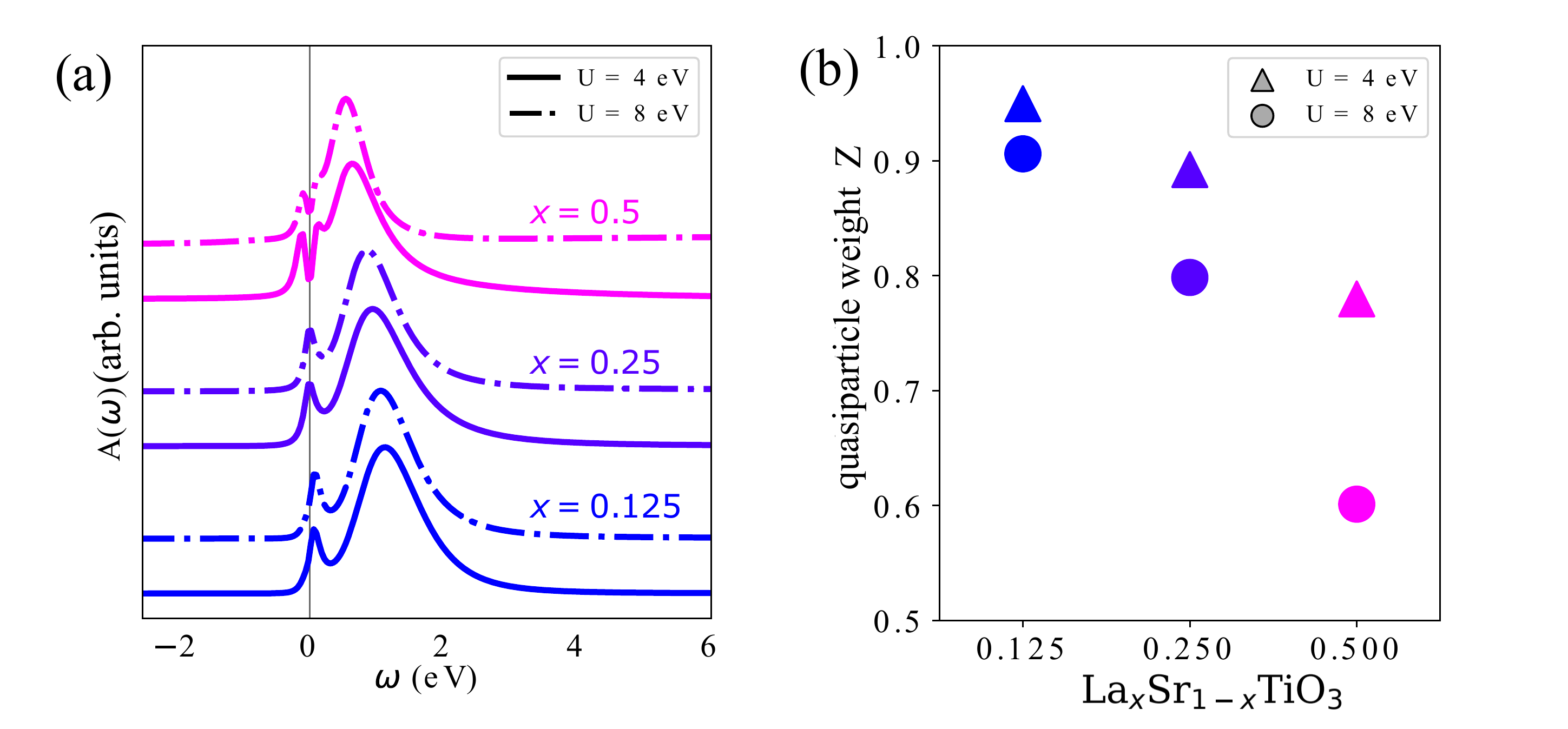}
\caption{(Color online) (a) Ti-\tgg DMFT spectral functions and  (b) corresponding quasiparticle weights for three different concentrations $x$ of La$_x$Sr$_{1-x}$TiO$_3$ for a $40$-atom unit cell and two different values of $U$. Spectral functions in (a) are shifted on the $y$-axis for clarity}
\label{fig:la_doping}
\end{figure}

As we mentioned in the Introduction, the question of how different doping sources affect the detailed electronic properties is an important one, with earlier studies suggesting that, while \ovv doping significantly alters the band structure as we have seen here, La$_{\rm Sr}$ causes only a rigid shift of the bands.~\cite{sarma98,ishida08} Here, we test whether the DFT+DMFT treatment is consistent with these earlier findings by calculating the behavior when a small concentration of Sr ions is replaced by La.

Fig.~\ref{fig:la_bands} shows the evolution of the DFT La$_x$Sr$_{1-x}$TiO$_3$ bandstructure for three different values of $x$ corresponding to La concentrations of less than or equal to half. These results confirm that the main features of the \STOO bandstructure are robust against doping. The only, yet rather important, difference between the three cases is the increasing partial filling of the Ti-\tgg bands with increasing La concentration. 

The DFT+DMFT spectral functions of the Ti-\tgg bands for these three La concentrations, shown in Fig.~\ref{fig:la_doping}(a), confirm the metallicity of the system. Their respective integrals up to $\omega=0$ yield the electron occupation of the corresponding orbitals, and in our simulations all have values consistent with a homogeneous distribution of the excess charge of one electron per La ion into the Ti bands. Increasing the strength of the interaction parameter from $U=4$\,eV to $U=8$\,eV does not have a noticeable effect on $A(\omega)$.
However, the corresponding quasiparticle weights, $Z$, (Fig.~\ref{fig:la_doping}(b)) are reduced from the ``uncorrelated'' value of $Z=1$, indicating some degree of electronic correlation in the corresponding bands. $Z$ decreases, and becomes more sensitive to the value of $U$, as the concentration of La increases, consistent with the vicinity to the Mott insulating state at $x=1$. 
Thus, our results confirm the validity of the rigid-band picture under La$_{\rm Sr}$ substitution, while at the same time indicating moderate correlation effects, increasing with La concentration.

\section{Summary and Conclusions}
\label{Conclusions}

We have presented a DFT+DMFT description of the long-discussed problem of oxygen vacancies in \STO, focusing on a controlled and systematic treatment of electronic correlations on the vacancy site.
This is achieved by using a minimal correlated subspace, which consists of the low energy Ti-\tgg orbitals plus an additional Wannier function located on the vacancy site. 
Our study reveals a strong influence of \uovv on the overall electronic structure of the system. 
Furthermore, our cRPA calculations support the importance of electronic correlations on the \ovv site, yielding a \uovv close to $60\%$ of \uti. 

In particular, we find a transition from a doubly-occupied (\uov$=$0) to a singly-occupied \ovv state (\uov$=$\uti), accompanied by a charge transfer of one electron from the defect state into the conduction band. 
The latter state is equivalent to the scenario proposed by Lin and Demkov based on a minimal model of the vacancy levels,~\cite{PhysRevLett.111.217601} and consistent with the spin-polarized GGA+${\cal U}$ results of Hou and Terakura,~\cite{hou10} but without the need for an artifical spin-symmetry breaking. It can in principle reconcile apparently contradictory experimental observations of low temperature metallic conductivity in combination with localized in-gap states.~\cite{moos97,ohtomo_hwang07,kan2005,yamada2009}
Our results are also consistent with a recent DFT+DMFT study of oxygen vacancies at the \STOO surface,~\cite{Lechermann_et_al:2016} which also found metallicity in combination with a split-off spectral feature inside the gap.

Analysis of the effect of the distance between two Ti ions separated by an \ovv on the nature of the \ovv charge leads to two important conclusions. First, different treatments of electronic interactions at the DFT(+${\cal U}$) level lead to different Ti-\ov-Ti distances. This in turn affects the relative \ovv and Ti-\tgg energy levels, with smaller Ti-\ov-Ti distances favouring localization of the charge on the \ovv and larger distances promoting partial filling of the Ti orbitals with one of the electrons released by the missing oxygen. Second, this picture is largely maintained after a DFT+DMFT analysis, with the \ovv electron occupation also depending on \uov. 
Future DFT+DMFT calculations considering full charge self-consistency and structural relaxations could provide further insight as to which Ti-\ov-Ti distance and vacancy site occupation is indeed  energetically preferred.

Lastly, we have compared  \ov-doped \STOO with the case of La$_\text{Sr}$ substitution, which represents another common way of electron doping the system. Our DFT+DMFT results, in line with previous DFT findings, show that La$_x$Sr$_{1-x}$TiO$_3$ is much closer to the {\it trivial} doping case, with no significant change in the bandstructure other than an increasing uniform filling of the Ti-\tgg bands by the extra electrons donated by the La cations. This demonstrates that different routes to electron-doping in \STOO are certainly not equivalent, and can lead to rather different electronic structures.

Finally, we mention that the method that we have demonstrated for controlling and analyzing the strength of electronic correlation on a vacancy defect state within the DFT+DMFT formalism is applicable to other defects beyond the oxygen vacancies studied here. Our finding that the electronic nature of the vacancy state depends strongly on the strength of the local interaction, \uov, highlights the need for such a methodology. We hope that, in addition to contributing to the ongoing debate about the case of doped \STO, our work will motivate similar studies on other materials in which correlated defects might play an important role.



\begin{acknowledgments}

We thank Peitao Liu and Cesare Franchini for fruitful discussions and technical help with the cRPA calculations.
This work was supported by the Swiss National Science Foundation through NCCR-MARVEL and by the K\"orber Foundation.
Calculations have been performed on the cluster ``Piz Daint'', hosted by the Swiss National Supercomputing Centre and supported under project IDs s889 (User Lab) and mr26 (MARVEL), and the ``Euler'' cluster of ETH Zurich. \\
\end{acknowledgments}

\bibliography{references}

\end{document}